%% file: main.tex
\documentclass{INTERSPEECH2023}

\interspeechcameraready

\usepackage{graphicx}
\usepackage{xcolor}
\usepackage{csquotes}
\usepackage[T1]{fontenc}
\usepackage{booktabs}
\usepackage{multirow}
\usepackage{multicol}
\usepackage{arydshln}

\graphicspath{./figures}

\title{Improving grapheme-to-phoneme conversion by learning pronunciations from speech recordings}

\name{Manuel Sam Ribeiro, Giulia Comini, Jaime Lorenzo-Trueba}

\address{Amazon Alexa, TTS Research}
\email{\{manuerib, gcomini, truebaj\}@amazon.com}

\begin{document}

\maketitle
\begin{abstract}
The Grapheme-to-Phoneme (G2P) task aims to convert orthographic input into a discrete phonetic representation.
G2P conversion is beneficial to various speech processing applications, such as text-to-speech and speech recognition. However, these tend to rely on manually-annotated pronunciation dictionaries, which are often time-consuming and costly to acquire. In this paper, we propose a method to improve the G2P conversion task by learning pronunciation examples from audio recordings. Our approach bootstraps a G2P with a small set of annotated examples. The G2P model is used to train a multilingual phone recognition system, which then decodes speech recordings with a phonetic representation. Given hypothesized phoneme labels, we learn pronunciation dictionaries for out-of-vocabulary words, and we use those to re-train the G2P system. Results indicate that our approach consistently improves the phone error rate of G2P systems across languages and amount of available data.
\end{abstract}

\input{./content/body.tex}

\bibliographystyle{IEEEtran}
\bibliography{references}

\end{document}

%% file: content/body.tex
\section{Introduction}

The Grapheme-to-Phoneme (G2P) task aims to convert orthographic input, a sequence of \emph{graphemes}, into a discrete phonetic representation, a sequence of \emph{phonemes}.
The ability to automatically convert graphemes into phonemes benefits speech processing systems such as Text-to-Speech (TTS) or Automatic Speech Recognition (ASR) by hypothesizing pronunciations for out-of-vocabulary words and generalizing beyond a finite, potentially small, manually-annotated pronunciation dictionary.

Specifically for TTS, systems often rely on large pronunciation dictionaries to control the phonetic output of generated speech samples. 
These dictionaries are time-consuming and challenging to acquire, as they require expert knowledge of both target language and phonetic transcription conventions. 
This is especially relevant for under-resourced or endangered languages. 
In the absence of phonetic knowledge, TTS models can be trained directly on graphemes \cite{shen2018natural} or some form of learned representations \cite{siuzdak2022wavthruvec}.
Such inputs, however, are problematic for languages with irregular orthography \cite{taylor2019analysis} and offer limitations when controlling or correcting the pronunciation of trained and deployed models \cite{taylor2019analysis}.
Additionally, phoneme-based TTS systems tend to outperform grapheme-based systems, provided that there is enough high-quality annotations, either through hand-crafted pronunciation dictionaries or G2P systems \cite{fong2019comparison}.

Traditional data-driven approaches to the G2P task used decision trees \cite{black1998issues}, hidden Markov models \cite{taylor2005hidden}, joint n-gram models \cite{galescu2002pronunciation}, weighted finite-state transducers \cite{novak2012wfst}, or neural networks \cite{damper1998comparison}. More recently, various forms of neural networks have been the default approach to G2P conversion, such as LSTMs with or without attention mechanisms \cite{toshniwal2016jointly, rao2015grapheme}, or transformer-based architectures \cite{yolchuyeva2020transformer}.
Modern neural network architectures easily outperform traditional data-driven systems \cite{yolchuyeva2020transformer, milde2017multitask, peters2017massively}.

Recent studies, therefore, shifted their focus towards unified multilingual G2P systems that aim to reduce the dissimilarities between languages \cite{peters2017massively, sokolov2020neural, elsaadany2020grapheme, yu2020multilingual, zhu2022byt5}.
Multilingual systems can also benefit under-resourced languages through cross-lingual knowledge transfer.
These systems aim to reduce the dependency on large pronunciation dictionaries to quickly scale to new languages \cite{sokolov2020neural, yu2020multilingual} or dialects \cite{engelhart2021grapheme}.
Alternative approaches to low-resource G2P models use unsupervised text-based pre-training \cite{dong2022neural}, or various forms of text-based data augmentation \cite{yu2020ensemble, hauer2020low, hammond2021data}.
Related work for automatic pronunciation learning proposed to acquire knowledge for new languages through audio examples in a zero-shot scenario.
This involves learning a new language's phonetic inventory \cite{zelasko2022discovering}, or the automatic labelling of pronunciation using universal phone recognition \cite{li2020universal, klejch2021deciphering} for downstream tasks such as ASR.
Specifically for G2Ps, related work also proposed to leverage audio data to iterate, revise, or complement a G2P system initialized on a small set of annotated materials \cite{goel2010approaches, route2019multimodal, aquino2019g2p}.

In this paper, we propose a method to \emph{improve Grapheme-to-Phoneme models by learning pronunciation examples for out-of-vocabulary words from audio recordings}.
Our approach builds on recent studies that use multilingual transformer-based G2P models for cross-lingual knowledge transfer \cite{sokolov2020neural, yu2020multilingual}, and automatic pronunciation learning from audio using phone recognition \cite{li2020universal, klejch2021deciphering}.
We describe our system in Section \ref{sec:method}, while in Section \ref{sec:experiments} we provide experimental evidence that our systems are scalable and effective in high- and low-resource scenarios.

\section{Method}
\label{sec:method}

\begin{figure}[t]
  \centering
  \includegraphics[width=\linewidth]{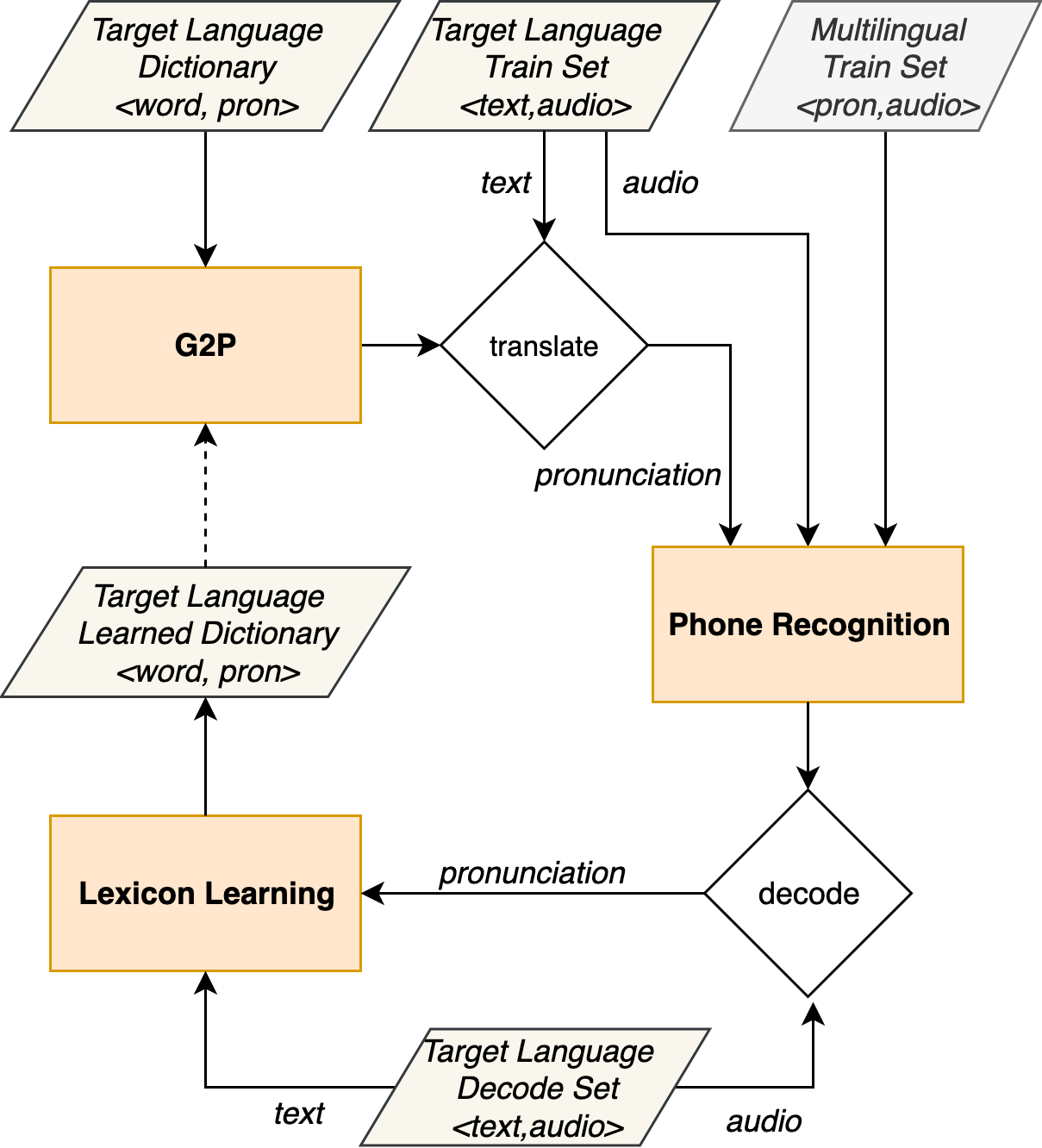}
  \caption{Approach to pronunciation learning from audio to improve low-resource G2Ps.}
  \label{fig:diagram}
\end{figure}

Figure \ref{fig:diagram} illustrates our approach to improve low-resource G2Ps by automatically learning novel pronunciations from speech recordings.
For an unseen target language, we assume that we have a hand-crafted pronunciation dictionary, consisting of <word, pronunciation> pairs, and a speech corpus consisting of <text, audio> pairs, typically used for TTS acoustic modelling.
We begin by training a baseline G2P model on the available pronunciation dictionary.
We then use the baseline G2P model to hypothesize pronunciations for the vocabulary in the target language speech corpus.
We train a Phone Recognizer using <pronunciation, audio> pairs from multilingual data, augmented with <pronunciation, audio> pairs in the target language.
We then decode the audio data in the target language. The decode set in the target language may or may not be the same as the train set.
Because the phone recognizer operates at the sentence level, we do not have knowledge of word boundaries.
The final lexicon learning step aims to align the decoded pronunciation sequences to observed character sequences for word boundary discovery and word-level lexicon learning.
In the following sections, we describe the architecture of learnable components of the proposed pipeline.

\subsection{Grapheme-to-phoneme conversion}

Our approach to G2P conversion is based on a transformer \cite{vaswani2017attention} encoder-decoder architecture, implemented with OpenNMT \cite{klein2017opennmt}.
The encoder and decoder each use 6 layers with 8 self-attention heads.
The hidden transformer feed-forward is set to 2048 nodes, while the embedding size is set to 512.
We set a dropout rate of 0.1 throughout the model.
The model is optimized with \texttt{adam}, using the \texttt{noam} learning rate scheduler with 8000 warmup steps \cite{vaswani2017attention}.
The G2P model operates at the word level and inputs a sequence of characters (graphemes), with a prepended language tag.
The output phoneme set is defined using the X-SAMPA \cite{wells1995computer} phonetic alphabet.
We pre-train the G2P model on a multilingual pronunciation corpus.
We then fine-tune the pre-trained multilingual model for a maximum of 20k steps on available <word, pronunciation> pairs in an unseen target language.
For experimental purposes, we ensure that the multilingual model does not have any knowledge of words that occur in unseen languages.

\subsection{Phone recognition}

Phone recognition aims to annotate speech recordings with the corresponding phoneme sequence.
We generate a pronunciation dictionary for the target language \enquote{Train Set} using the baseline G2P system.
This data is pooled with a large multilingual speech dataset, for which pronunciations are available from large hand-crafted dictionaries.
We implement our phone recognition system using Kaldi \cite{povey2011kaldi}.
We extract Mel Frequency Cepstral Coefficients (MFCCs) from the audio data, which are used to initialize monophone and triphone HMM-GMM models.
We then apply Linear Discriminant Analysis (LDA) and a maximum Likelihood Linear Transform (MLLT), which is followed by Speaker Adaptive Training with feature-space MLLR (fMLLR, \cite{rath2013improved}).
We use the features and alignments after this stage to train a time-delay neural network (TDNN, \cite{peddinti2015time}) acoustic model.
We do not explicitly provide the acoustic model with language identifiers.
The expectation is that the acoustic model will learn to generalize from acoustic realizations across multiple languages, overcoming the potentially noisy labels given by the G2P system for the target language.
This is similar to recent approaches to Universal Phone Recognition \cite{li2020universal, klejch2021deciphering}.
When decoding speech samples in the target language, we use a 5gram language model learned on the phone-level generated transcripts for the \enquote{Train Set}.
The language model enforces the system to decode plausible pronunciations restricted to the target language's phonotactics, guided by generated G2P output.
The acoustic model aims to hypothesize labels informed by audio samples, guided by universal phonetic knowledge given by multilingual speech examples.

\subsection{Lexicon learning}

Because we are decoding speech segments at the sentence level, the output of the phone recognition system is a sequence of phonemes without word boundaries.
We address this by force-aligning the decoded phoneme sequences to the observed grapheme sequences.
We define hidden Markov models (HMMs) for each word in the \enquote{Decode Set} vocabulary.
The HMMs follow a left-to-right topology with skip connections, with each state corresponding to a grapheme in the word.
The probability of each state aligning to a given phoneme is given by a discrete probability distribution over the phoneme space.
These distributions are initialized to a uniform distribution and tied across states/graphemes.
For each sentence in the \enquote{Decode Set}, we concatenate the word-level HMMs to form a large sentence-level HMM.
Using Viterbi, we find the most likely path through the sentence graph, given the decoded pronunciation sequence.
HMM models are implemented and optimized using \texttt{pomegranate} \cite{schreiber2018pomegranate}.

Given hypothesized word boundaries, we can generate learned pronunciation dictionaries by collecting decoded pronunciations for each word in the \enquote{Decode Set} vocabulary.
Because words can occur multiple times, we define an acceptance threshold $k$, denoting the number of times we must observe a phoneme sequence paired with any given word.
That is, in order to allow a <word, pronunciation> pair to be included in the learned dictionary, we must have decoded the pronunciation for that word independently at least $k$ times.
Higher $k$ discards infrequently decoded pronunciations and implies a small learned dictionary of high-frequency words, but higher confidence in the learned pronunciations.

\begin{table*}[th]
  \caption{Phone Error Rate (PER) for pronunciation dictionaries learned at various thresholds $k$ using a seed set of 500 words (\emph{PER Learned}) and equivalent dictionaries generated by a baseline G2P system (\emph{PER G2P}). Results are presented for the average across 5 languages and separately for English. Columns \enquote{Better}, \enquote{Worse}, and \enquote{Same} indicate the number of words that are better, worse, or the same with respect to the baseline G2P for the English learned dictionaries.}
  \label{tab:learned_lexica}
  \resizebox{\linewidth}{!}{%
\begin{tabular}{@{}c|cc|cccccc@{}}

     & \multicolumn{2}{c|}{Average (5 languages)}  & \multicolumn{6}{c}{English} \\ \toprule
$k$  & PER (Learned)      & PER (G2P)         & PER (Learned)     & PER (G2P)   & Num Words & Better           & Worse         & Same                \\ \midrule
1    & 12.99\%            & \textbf{12.53\%}  & \textbf{15.32\%}  & 17.31\%     & 26557     & 4360   (16.42\%) & 2431 (9.15\%) & 19766 (74.43\%)     \\
2    & \textbf{9.27\%}    & 10.25\%           & \textbf{11.13\%}  & 13.45\%     & 12271     & 1328 (10.82\%)   & 359 (2.93\%)  & 10584 (86.25\%)     \\
4    & \textbf{7.86\%}    & 8.55\%            & \textbf{9.15\%}   & 10.60\%     & 5943      & 364 (6.12\%)     & 90 (1.51\%)   & 5489 (92.36\%)      \\
6    & \textbf{7.33\%}    & 7.80\%            & \textbf{8.57\%}   & 9.38\%      & 3962      & 153 (3.86\%)     & 56 (1.41\%)   & 3753 (94.72\%)      \\
8    & \textbf{6.93\%}    & 7.23\%            & \textbf{8.19\%}   & 8.64\%      & 3009      & 72 (2.39\%)      & 40 (1.33\%)   & 2897 (96.28\%)      \\ \bottomrule

\end{tabular}%
}
\end{table*}

\section{Experiments}
\label{sec:experiments}

We evaluate our method on five languages: English, French, Danish, Polish, and Turkish.
We begin by training a multilingual G2P model for 1M steps on a training set of 4.5M <word, pronunciation> pairs, pooled across 17 languages.
This model is then fine-tuned independently for each of the target languages on a set of seed word pronunciations.
We consider initially a low-resource scenario using a seed set of 500 manually-revised words.
We select the most frequent 500 words, based on word counts from roughly 2M sentences, extracted from the multilingual C4 corpus \cite{xue-etal-2021-mt5}.
We use the baseline fine-tuned G2P system to provide pronunciations for a set of audio recordings in the target language.
This \enquote{Train/Decode Set} consists of single-speaker studio-quality recordings used for TTS model training.
Depending on the language, the amount of available data ranges from 9 to 27 hours.
The target-language data is complemented with approximately 165 hours of multilingual speech data, pooled from ~6k speakers across 16 languages.
The size of the \enquote{Multilingual Train Set} differs per language, as we exclude target language data from each iteration.
After lexicon learning, we pool the learned pronunciations with the manually-revised seed set and fine-tune the pre-trained multilingual G2P model.
We evaluate the G2P systems on test sets consisting of word tokens unseen at training time, and measure results in terms of Phone Error Rate (PER) and Word Error Rate (WER).

\begin{table}[]
\centering
  \caption{Phone and Word error rates (PER/WER) for baseline and learned G2P systems at various lexicon learning thresholds $k$. PER reduction is computed relative to the baseline G2P. Results are averaged across 5 languages.}
  \label{tab:results_by_k}
  \resizebox{\linewidth}{!}{%
\begin{tabular}{@{}ccccc@{}}
\toprule
System                     & $k$ & PER   & WER     & PER Rel. Reduction  \\ \midrule
Baseline                   & -   & 13.02\% & 50.16\% & -               \\ \hdashline\noalign{\vskip 0.9ex}
\multirow{5}{*}{Learned} & 1   & \textbf{10.64\%} & \textbf{45.39\%} & \textbf{-18.32\%}        \\
                           & 2   & 11.02\% & 46.06\% & -15.36\%        \\
                           & 4   & 11.47\% & 46.50\% & -11.94\%        \\
                           & 6   & 11.69\% & 47.03\% & -10.23\%        \\
                           & 8   & 11.86\% & 47.53\% & -8.89\%         \\ \bottomrule
\end{tabular}%
}
\end{table}

\subsection{Lexicon learning}

We investigate the impact of the threshold $k$ on the learned pronunciation dictionaries.
Table \ref{tab:learned_lexica} shows error rates for the learned dictionaries averaged across the 5 languages, with additional details for English.
We include results for equivalent G2P-generated dictionaries, which contain the same words as the corresponding learned dictionaries, but with pronunciations generated by the baseline G2P.
As expected, as we increase $k$, the amount of words allowed in the dictionaries decreases, and so does the overall error rate of the dictionary.
When considering the average results, we observe that the dictionaries learned at $k=1$ underperform when compared with equivalent G2P-generated dictionaries.
However, this is not the case for English, where we observe improvements across all values of $k$.
This might be due to amount of available data for the \enquote{Train/Decode Sets}. 
The English set has 27 hours of speech data, whereas the underperforming languages at $k=1$, Polish and Turkish, have 9 and 14 hours, respectively.
When considering the distribution of the learned pronunciations (English example in Table \ref{tab:learned_lexica}) , we observe that most decoded pronunciations are the same as those given by the baseline G2P system.
However, when the pronunciations are not the same, the phone recognition system has a positive impact over the G2P-generated pronunciations. 

\begin{table}[]
  \caption{Phone Error Rate (PER) for baseline and learned G2P systems at $k=1$ and $k=2$ for five languages. Figures in parenthesis indicate the PER reduction relative to the corresponding baseline system.}
  \label{tab:results_by_language}
  \resizebox{\linewidth}{!}{%
\begin{tabular}{@{}cccc@{}}
\toprule
Language & Baseline & $k=1$                & $k=2$                \\ \midrule
English  & 19.20\%  & \textbf{16.42\%} (-14.48\%) & 17.01\% (-11.38\%) \\ 
Danish   & 20.57\%  & \textbf{16.94\%} (-17.67\%) & 17.76\% (-13.66\%) \\
French   & 12.13\%  & \textbf{8.09\%} (-33.28\%)  & 8.86\% (-26.93\%)  \\
Turkish  & 9.27\%   & 8.44\% (-8.90\%)  & \textbf{8.41\%} (-9.28\%)   \\
Polish   & 3.95\%   & 3.30\% (-16.58\%) & \textbf{3.07\%} (-22.28\%)  \\ \hdashline\noalign{\vskip 0.9ex}
\emph{Average}  & \emph{13.02\%}  & \emph{\textbf{10.64\%} (-18.32\%)}   & \emph{11.02\% (-15.37\%)}  \\ \bottomrule
\end{tabular}%
}
\end{table}

\subsection{Grapheme-to-Phoneme conversion}

We investigate the impact of the learned dictionaries at various thresholds $k$ on the fine-tuned G2P systems.
We pool each learned dictionary with the manually-annotated seed set of 500 words and re-train the multilingual G2P system.
Table \ref{tab:results_by_k} shows results for all values of $k$, averaged across the 5 languages.
We observe a positive impact across all tested values of $k$, with the best results occurring when $k=1$.
This is an interesting observation, as we have noted that, on average, learned dictionaries at $k=1$ underperform when compared with corresponding G2P-generated dictionaries (Table \ref{tab:learned_lexica}).
These figures suggest that, in a low-resource scenario, the G2P benefits from additional training data, even if containing a higher error rate. 
In other words, quantity appears to be preferable to quality.
We do note also that the G2P likely benefits from the large number of words that were already predicted correctly by the G2P system. 
Rather than just correcting mistakes from the G2P, the phone recognizer also validates correct pronunciation examples.
Table \ref{tab:results_by_language} shows  results for each of the evaluated languages at $k=1$ and $k=2$.
Turkish and Polish achieve the best results at $k=2$.
We hypothesize that this might be due to the already low error rates of the G2P for these languages.
Turkish and Polish have more straightforward grapheme-to-phoneme correspondence, when compared with the other languages evaluated.
It might be that, for such cases, higher thresholds are more suitable, leading to smaller, but higher quality dictionaries.

\subsection{Amount of seed data}
\begin{figure}[t]
  \centering
  \includegraphics[width=\linewidth]{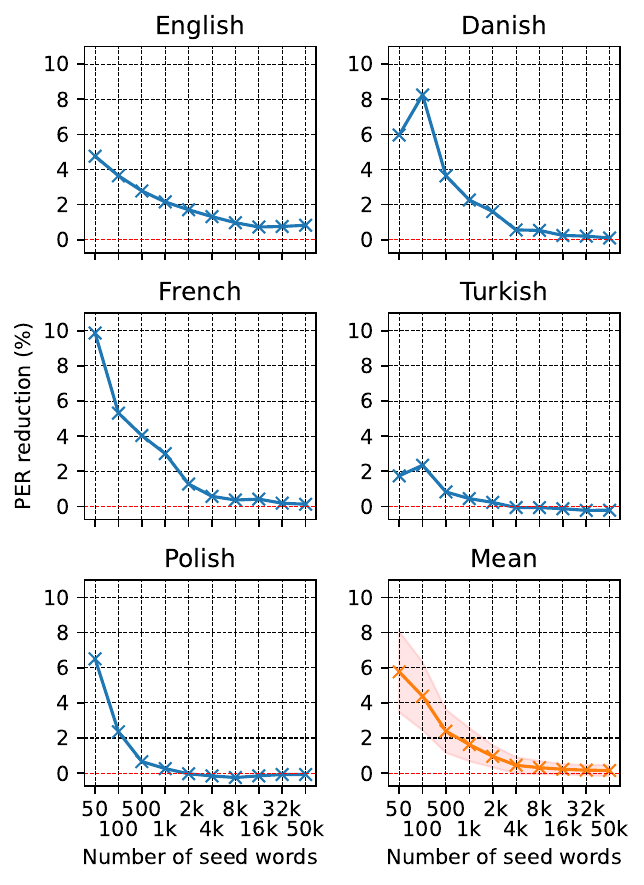}
  \caption{Phone Error Rate (PER) reduction between baseline and learned G2P system across varying number of seed words. Results are presented for 5 languages and their mean.}
  \label{fig:amount_of_data}
\end{figure}

Thus far, we have presented results for low-resource G2Ps using a seed set of 500 annotated words.
We investigate the behaviour of our approach as we vary the amount of seed data.
The pipeline is identical to that described in Section \ref{sec:method}, and we change only the size of the manually-annotated pronunciation dictionaries for the target language.
We define the dictionaries such that each incremental word set is a superset of the one immediately before.
Figure \ref{fig:amount_of_data} illustates results for the 5 languages and their average.
For simplicity, we visualize the absolute PER difference between the baseline and the learned system at $k=1$.
We observe that our method has the highest impact on low-resource scenarios, when 1000 words or less are available.
For these systems, we reduce the PER by 2-6\%, on average across all languages.
The biggest improvements occur with French, Danish and Polish when only with 50 annotated words are available.
Our approach reduces PER by 6-10\% absolute over the baseline system.
For medium and high-resource scenarios (more than 2000 words), results show variable improvements across languages.
Languages such as Polish and Turkish observe a deterioration of PER over the baseline, whereas languages such as English, Danish or French observe an improvement. These changes, however, are small ({\raise.17ex\hbox{$\scriptstyle\mathtt{\sim}$}}1\% PER).

\begin{table}[th]
  \caption{Phone Error Rate (PER) over multiple iterations of pronunciation learning for English and Danish using a seed set of 100 and 500 words at $k=1$. The columns \enquote{Rel. Red.} denote the PER reduction relative to the previous iteration.}
  \label{tab:results_by_iterations}
\resizebox{\linewidth}{!}{%
\begin{tabular}{@{}cc|cc|cc@{}}
\toprule
\multirow{2}{*}{Language} & \multirow{2}{*}{System}  & \multicolumn{2}{c}{100 seed words}                 & \multicolumn{2}{c}{500 seed words}    \\  \cmidrule(){3-6}
  &   & PER     & Rel. Red. & PER     & Rel. Red.          \\ \midrule
\multirow{6}{*}{English} & Baseline       & 22.26\%   & -           & 22.02\%   & -            \\  \cdashline{2-6}\noalign{\vskip 0.9ex}
& Iter 1         & 18.44\%   & -17.14\%    & 17.62\%   & -11.97\%     \\
& Iter 2         & 17.71\%   & -3.96\%     & 17.16\%   & -2.61\%      \\
& Iter 3         & 17.71\%   & 0.00\%      & 16.97\%   & -1.14\%      \\
& Iter 4         & 17.69\%   & -0.14\% & \textbf{16.87\%}   & -0.59\%      \\
& Iter 5         & \textbf{17.59\%}  & -0.54\%         & 17.03\%   & 0.95\%          \\ \midrule
\multirow{6}{*}{Danish}  & Baseline       & 39.60\%   & -           & 20.63\%    & -           \\ \cdashline{2-6}\noalign{\vskip 0.9ex}
& Iter 1         & 30.56\%   & -22.81\%    & \textbf{17.29\%}    & -16.19\%    \\
& Iter 2         & 30.26\%   & -1.00\%     & 17.31\%    & 0.12\%      \\
& Iter 3         & 29.81\%   & -1.50\%     & 17.35\%    & 0.23\%      \\
& Iter 4         & \textbf{29.56\%}   & -0.82\%     & 17.46\%    & 0.63\%      \\
& Iter 5         & 29.56\%   & 0.00\%      & 17.43\%    & -0.20\%     \\ \bottomrule
\end{tabular}%
}
\end{table}

\subsection{Iterative self-training}

Given the promising results after a single pass of our pipeline, we investigate additional improvements as we iterate in a type of \enquote{self-training} scenario.
For each iteration, we choose a G2P checkpoint based on the phone error rate measured on a small validation set.
This checkpoint is then used to annotate the audio data for phone recognition.
All systems are trained from scratch for each iteration using the same audio data and a constant $k=1$.
This is a slightly different setup from earlier experiments, which is why baseline figures differ.
Results are presented in Table \ref{tab:results_by_iterations} for the evaluation sets.
Although the impact of the learned pronunciation dictionaries decreases as we iterate, we always observe improvements by self-training.
For example, in the English system with a seed set of 100 words, we reduce PER relative to the baseline from 17.14\% (iteration 1) to 20.96\% (iteration 5).

\section{Conclusion and future work}

We have proposed an approach to improve grapheme-to-phoneme models by learning pronunciations from speech recordings.
We showed evidence that our method consistently improves G2P systems for low-resource scenarios across 5 different languages.
Results indicate that pronunciation dictionaries learned with a low threshold lead to the best results.
This suggests that the quantity of words tends to be preferred over their quality.
The impact of the pronunciation learning system is higher for low-resource scenarios, when the G2P system is more prone to errors.
We additionally observe that iterating over the pipeline leads to increase performance for the G2P system.

For future work, we aim to investigate methods for automatic pronunciation learning in high-resource scenarios.
Our results suggest that the impact of our approach is small when a large amount of data is available.
We aim to investigate solutions to learn novel pronunciations for infrequent words with irregular grapheme-to-phoneme correspondence, such as proper names, loan words, or domain-specific tokens.
Our experimental evidence relied on target language audio data that was restricted to a single-speaker set of recordings.
Further work should explore the impact of the type or amount of audio data in the target language.
Using larger multi-speaker speech recordings in the target language might lead to improved results, particularly for the low-resource scenarios.
Additionally, we might achieve improved performance with more robust speech recognition systems, perhaps pre-trained in a self-supervised fashion.